\documentclass[pdflatex,sn-mathphys-num]{sn-jnl}

\usepackage[T1]{fontenc}
\usepackage[utf8]{inputenc}
\usepackage{graphicx}
\usepackage{multirow}
\usepackage{amsmath,amssymb,amsfonts}
\usepackage{amsthm}
\usepackage{mathrsfs}
\usepackage[title]{appendix}
\usepackage{xcolor}
\usepackage{textcomp}
\usepackage{manyfoot}
\usepackage{booktabs}
\usepackage{listings}
\usepackage{optidef}
\usepackage{subcaption}
\usepackage{times}
\usepackage{enumerate}
\usepackage{latexsym}
\usepackage{pdfpages}
\usepackage{adjustbox}
\usepackage{enumitem}
\usepackage{array, bigdelim, makecell}
\usepackage{url}
\usepackage{hyperref}
\usepackage{setspace}
\usepackage[lined,boxed,commentsnumbered,procnumbered,ruled]{algorithm2e}
\usepackage{titlesec}
\usepackage{tcolorbox}
\tcbuselibrary{skins,breakable}
\usepackage{algpseudocode}
\usepackage{microtype}

\titleformat{\paragraph}[runin]{\normalfont\bfseries}{\theparagraph}{1em}{}
\makeatletter
\def\@xfootnote[#1]{%
  \protected@xdef\@thefnmark{#1}%
  \@footnotemark\@footnotetext}
\makeatother
 \SetKwFunction{FRecurs}{FnRecursive}%

\begin{document}

\title{MADS: A Multiview Acoustic Descriptor Set Beyond Standard Spectral Summaries}

\author*[1]{\fnm{Utsab} \sur{Ghosh}}
\author[1]{\fnm{Roshni} \sur{Chakraborty}}

\affil[1]{\orgname{ABV-Indian Institute of Information Technology and Management, Gwalior}, \country{India}}

\abstract{Dominant audio classification pipelines rely either on compact handcrafted summaries or on fixed time-frequency frontends such as log-mel representations prior to deep modeling. While highly successful, these representations do not explicitly expose the physical dynamics of the underlying sound-generating event. We introduce \emph{MADS} (\textbf{M}ulti-view \textbf{A}coustic \textbf{D}escriptor \textbf{S}et), a compact 19-dimensional physics-informed descriptor set designed to capture complementary spectral, temporal, mechanical, and stochastic structure in audio signals. Rather than treating sound only as a spectral pattern, \emph{MADS} encodes properties related to excitation, damping, periodicity, impulsiveness, and structural consistency within a unified multi-view representation.

We evaluate \emph{MADS} using standard classical machine learning models on \textit{ESC-10}, \textit{ESC-50}, and \textit{MSoS}, and compare it against two conventional handcrafted baselines: a compact 26D MFCC-based baseline and an expanded 38D spectral-summary baseline. Across \textit{ESC-10} and \textit{ESC-50}, \emph{MADS} achieves the strongest peak results overall, reaching \textbf{81.00\%} and \textbf{52.78\%}, respectively, while using roughly half the dimensionality of the 38D baseline. On \textit{MSoS}, \emph{MADS} again delivers the strongest top-end performance, reaching \textbf{67.48\%}. These results establish \emph{MADS} not merely as a competitive standalone descriptor set, but as the foundational descriptor layer of a broader acoustically grounded representation program for future frame-level and deep-learning-compatible audio modeling.}


\maketitle

\section{Introduction}
Sound classification is important in applications such as urban monitoring, machine condition awareness, assistive sensing, surveillance, and autonomous systems, where audio can provide situational awareness even when visual information is unreliable. Large-scale pretrained audio models such as PANNs and transformer-based systems have substantially improved sound classification by learning strong representations from time-frequency inputs \cite{kong2020panns,gong2021ast,chen2022hts}. More recent models such as BEATs and CLAP extend this trend through self-supervised acoustic tokenization and audio-language supervision, respectively \cite{chen2022beats,elizalde2022clap}. Although these models are highly effective, they remain largely data-driven, i.e., the waveform is converted into a time-frequency representation and processed as a high-dimensional pattern rather than as the output of a physical sounding process. As a result, acoustically distinct events such as \emph{wind}, \emph{steam-like turbulence}, and \emph{machinery hum} can occupy overlapping spectral regions even when their underlying excitation, dissipation, and temporal structure differ substantially.

\par A parallel line of research has relied on handcrafted descriptors such as Mel-frequency cepstral coefficients (MFCCs) \cite{davis1980comparison} and standard spectral and temporal summary features including spectral centroid, zero-crossing rate, spectral flatness, roll-off, and related statistics \cite{peeters2004large}. These descriptors remain attractive because they are compact, interpretable, and inexpensive to compute. Hybrid pipelines have also explored combining handcrafted descriptors with learned representations through late fusion of a handcrafted-feature classifier and a deep model \cite{fonseca2018simple}.

\par However, standard handcrafted descriptors are primarily designed as compact summaries of spectral shape and local variation, and are therefore not explicitly targeted at resolving event-level dynamic differences. In practice, they can describe where energy is concentrated and how the spectrum is shaped, but they are less expressive about how the event evolves over time, how energy is transferred, and how impulsive, damped, periodic, frictional, or mechanically unstable the source behavior may be.

\par At the same time, modern deep-learning systems for sound classification largely rely on log-mel representations of raw waveforms. Large-scale pretrained systems such as PANNs \cite{kong2020panns}, AST \cite{gong2021ast}, HTS-AT \cite{chen2022hts}, BEATs \cite{chen2022beats}, and related architectures all build on time-frequency representations that are either explicitly mel-based or closely tied to the same frontend assumptions. This design choice has been highly successful in practice, but it is not beyond question. Prior work has pointed out several limitations of the standard log-mel pipeline. The mel scale itself has been revised multiple times \cite{o1987speech,umesh1999fitting}, and the auditory evidence originally used to support it was later questioned \cite{greenwood1997mel}. More broadly, a perceptually motivated frequency scale may not always be ideal for tasks that require different spectral resolution profiles. These observations have motivated a broader line of research on replacing fixed mel-filterbanks with learnable neural frontends, including Gammatone-inspired frontends \cite{sainath2015learning}, sinc-based models such as SincNet \cite{ravanelli2018speaker}, and more recent learnable frontends such as LEAF \cite{zeghidourleaf}. In particular, LEAF showed that replacing standard mel-filterbanks and fixed compression with a learnable frontend can be beneficial in several audio tasks, suggesting that the dominant log-mel pipeline remains an open design space rather than a settled optimum \cite{zeghidourleaf}. However, even this line of work largely focuses on improving how spectral frontends are learned, rather than explicitly representing the underlying physical dynamics of the sound-generating event itself.

\par Motivated by this broader line of research, we propose \textit{MADS}, a 19-dimensional descriptor set for sound classification. Although audio classifiers observe digital pressure sequences, the underlying events are generated by physical processes such as impacts, oscillations, damping, friction, bursts, and turbulent energy transfer. To explicitly model these dynamics, \textit{MADS} combines complementary spectral, temporal, mechanical, and stochastic views of an audio sample. By unifying these four distinct perspectives, we construct a multi-view, physics-informed feature space that provides a compact, interpretable, and structured descriptor basis for representation. In this sense, the present work is foundational: our goal is not merely to validate a standalone compact representation, but to establish a descriptor layer that can support future frame-level and deep-learning-compatible extensions.

\par We evaluate the proposed 19-dimensional descriptor set on \textit{ESC-10} \cite{piczak2015esc}, \textit{ESC-50} \cite{piczak2015esc}, and \textit{MSoS} \cite{kroos2019msos} using a standard suite of classical machine learning classifiers, and compare it against two conventional handcrafted baselines: a compact 26D MFCC-based baseline and an expanded 38D baseline built from standard spectral summary descriptors. Across \textit{ESC-10} and \textit{ESC-50}, \textit{MADS} consistently outperforms the compact baseline and achieves the strongest peak results overall, reaching \textbf{81.00\%} on \textit{ESC-10} and \textbf{52.78\%} on \textit{ESC-50}, compared to the best baseline results of \textbf{79.45\%} and \textbf{51.78\%}, respectively, while using substantially fewer features than the 38D baseline. On \textit{MSoS}, \textit{MADS} again delivers the strongest top-end performance, reaching \textbf{67.48\%} with the voting ensemble, compared to the best baseline result of \textbf{66.04\%}. Taken together, these results position \textit{MADS} as both a practically useful standalone descriptor set and a foundational representation for future acoustically grounded audio modeling.
\footnote{Due to double Blind review protocol the code link to the github repo is not provided and will be provided on acceptance}

.

\section{Proposed Methodology}
In this Section, we initially discuss the Problem Definition followed by details of the proposed features, \textit{MADS}, i.e., Multi-View Acoustic Descriptor Set, in detail.
\subsection{Problem Statement}

\par Environmental sounds are often represented through spectral summaries that capture coarse timbre and energy distribution, but not necessarily the physical dynamics of the underlying source event \cite{gaver1993world}. As a result, different sound-producing mechanisms can exhibit similar spectral structure even when their excitation, dissipation, periodicity, or impulsiveness differ substantially.

\par We therefore ask whether a compact set of acoustically grounded descriptors can recover some of this missing physical structure and provide a stronger, more interpretable basis for environmental sound classification. Let \(x[n]\) denote an input waveform and \(y \in \mathcal{Y}\) its class label. While conventional front-ends construct a spectral representation \(\phi_{\mathrm{spec}}(x)\), our objective is to define a compact descriptor map
\[
\phi_{\mathrm{dyn}}(x) \in \mathbb{R}^{19},
\]
that preserves physically meaningful structure and provides an effective basis for predicting \(y\).

\subsection{Multi-View Acoustic Descriptor Set, \textit{MADS}}
While a digital audio signal is a dimensionless sequence of pressure values, it is the product of a complex dynamical system involving a physical radiator, a propagation medium, and stochastic energy exchange. To bridge the discriminative gap inherent in standard spectral summaries, we use four different lenses to look at the audio:
\begin{enumerate}
    \item \textbf{Mechanical Descriptors} : We use this term to denote descriptors derived from discrete velocity, acceleration, and energy, designed to characterize coupling, dissipation, and instability directly from the waveform. In this group we include \textit{Kinetic Energy Entropy}, \textit{Energy Phase Correlation}, \textit{Mechanical Efficiency}, and \textit{Mechanical Instability Index}. 

\item \textbf{Spectral Anchors} : These descriptors summarize the coarse spectral envelope, dominant resonance, and effective bandwidth of the signal using standard frequency-domain statistics \cite{peeters2004large,davis1980comparison}. In this group we include \textit{Spectral Anchor I--III}, \textit{Dissipation Limit}, \textit{Resonant Spread}, and \textit{Peak Mode}.

    \item \textbf{Temporal Dynamics} : These descriptors represent onset behavior, temporal roughness, and the density of salient excitation events over time \cite{peeters2004large}. In this group we include \textit{Attack Gradient}, \textit{Damping Coefficient}, \textit{Temporal Jitter}, and \textit{Pulse Density}.

    \item \textbf{Stochasticity \& Consistency} : These descriptors quantify how ordered, irregular, broadband, or periodically structured the signal is using flatness, kurtosis, zero-crossing variation, autocorrelation, and residual consistency measures \cite{peeters2004large}. In this group we include \textit{entropy flow}, \textit{Impact Peakedness}, \textit{ZCR Variability}, \textit{Periodicity Strength}, and \textit{PDE Residual}.
\end{enumerate}

\paragraph{Acoustic Mechanics Foundations}
To construct mechanically grounded descriptors from the waveform, we introduce a small set of discrete quantities. Let $x[n]$ denote the normalized audio waveform. We define an instantaneous velocity as the first-order difference
difference as shown in Equation \ref{eq:f1} and acceleration as the second-order difference as shown in in Equation \ref{eq:f2} next
\begin{equation}
v[n] = x[n] - x[n-1].
\label{eq:f1}
\end{equation}

\begin{equation}
a[n] = v[n] - v[n-1].
\label{eq:f2}
\end{equation}
These quantities are used as discrete proxies that allow us to characterize effort, dissipation, and temporal energy transfer directly from the signal. We further define an effective mass term \(m_{\mathrm{eff}}\) from the spectral centroid trajectory\cite{peeters2004large}. Let
\begin{equation}
c[t] = \frac{\sum_k f_k |X_k(t)|}{\sum_k |X_k(t)|}
\label{eq:f3}
\end{equation}
denote the spectral centroid at frame \(t\), i.e., the spectral center of gravity of the spectrum \cite{iso15938audio}. Motivated by this center-of-gravity interpretation, we use the centroid trajectory to construct an effective mass proxy. We then define

\begin{equation}
m_{\mathrm{eff}} = \frac{\mu_c + \sigma_c}{f_s},
\end{equation}
where, $\mu_c$ and $\sigma_c$ are the mean and standard deviation of the centroid trajectory, and $f_s$ is the sampling rate. This term acts as a heuristic inertia-like scaling factor for the discrete mechanics used below. Using $m_{\mathrm{eff}}$, we define three foundational energetic quantities as follows in Equation  \ref{eq:f4},\ref{eq:f5}, \ref{eq:f6}, which serve as the basis for the mechanical descriptors, discussed next.
\begin{equation}
E_k[n] = \frac{1}{2} m_{\mathrm{eff}} v[n]^2
\label{eq:f4}
\end{equation}
\begin{equation}
E_p[n] = \frac{1}{2} m_{\mathrm{eff}} x[n]^2
\label{eq:f5}
\end{equation}
\begin{equation}
W[n] = \big(m_{\mathrm{eff}} a[n]\big)\, v[n]
\label{eq:f6}
\end{equation}

These serve as the foundational values for Mechanical Energetic Descriptors

\subsubsection{Mechanical Energetics}
These descriptors model the waveform as a discrete mechanical system and quantify how energy is distributed, coupled, and dissipated over time.
\begin{itemize}
    \item \textbf{Kinetic Energy Entropy} ($H_{ke}$): This descriptor measures how concentrated or distributed the kinetic energy is over time. Sounds such as \emph{mouse click}, \emph{can opening}, and \emph{water drops} tend to produce a few dominant energetic events and therefore lower entropy, whereas \emph{vacuum cleaner}, \emph{washing machine}, and \emph{siren} exhibit more uniformly distributed energy and therefore higher entropy. We compute it as the Shannon entropy \cite{shannon1948mathematical} of the normalized kinetic energy sequence as shown next:
    \begin{equation}
    H_{ke} = -\sum_n \hat{E}_k[n]\log_2(\hat{E}_k[n] + \epsilon),
    \end{equation}
    where \(\hat{E}_k[n]\) denotes the normalized kinetic energy at sample \(n\), and \(\epsilon\) is a small constant for numerical stability.
    \begin{equation}
    \hat{E}_k[n] = \frac{E_k[n] + \epsilon}{\sum_n \left(E_k[n] + \epsilon\right)}.
    \end{equation}

    \item \textbf{Energy Phase Correlation} ($\rho_{kp}$): This descriptor measures the degree of coupling between kinetic and potential energy states. Higher values arise in impulsive or sharply structured events such as \emph{mouse click}, \emph{can opening}, and \emph{glass breaking}, where multiple energy components peak together, while lower values are more characteristic of smoother or sustained sources such as \emph{siren}, \emph{wind}, and \emph{engine}. We compute it as the Pearson correlation \cite{pearson1896vii} between the kinetic and potential energy sequences:
\begin{equation}
\rho_{kp} =
\frac{\mathrm{cov}(\tilde{E}_k,\tilde{E}_p)}
{\sigma_{\tilde{E}_k}\sigma_{\tilde{E}_p}}.
\end{equation}
Here, \(\tilde{E}_k\) and \(\tilde{E}_p\) denote the kinetic and potential energy sequences truncated to a common length, \(\sigma_{\tilde{E}_k}\) and \(\sigma_{\tilde{E}_p}\) denote the standard deviations of the truncated kinetic and potential energy sequences, respectively. If this correlation is numerically undefined, it is set to \(0\).

    \item \textbf{Mechanical Efficiency} ($\eta$): This descriptor quantifies how much kinetic energy is retained relative to the mean mechanical work. Signals such as \emph{siren}, \emph{thunderstorm}, and \emph{airplane} exhibit higher values because they retain energy in a relatively stable or sustained state, whereas \emph{can opening}, \emph{crackling fire}, and \emph{glass breaking} exhibit lower values because energy is dissipated more irregularly. We define it as a log-scaled ratio between mean kinetic energy and the absolute mean work:
\begin{equation}
\eta =
\log\!\left(
1 + \left|
\frac{\mathbb{E}[E_k[n]]}
{|\mathbb{E}[W[n]]| + \epsilon}
\right|
\right).
\end{equation}
Here, \(\mathbb{E}[E_k[n]]\) is the mean kinetic energy over the clip and \(|\mathbb{E}[W[n]]|\) denotes the absolute value of the meanof the work sequence over the clip.
 and \(\epsilon\) is a small constant for numerical stability.

    \item \textbf{Mechanical Instability Index} ($\Psi$): This descriptor measures higher-order instability in the signal by combining acceleration variation and work variability. It is higher for mechanically irregular or intermittent sounds such as \emph{clock alarm}, \emph{can opening}, \emph{glass breaking}, and \emph{crickets}, and lower for smoother or more stable sources such as \emph{cow}, \emph{church bells}, and \emph{siren}. We define it as
\begin{equation}
\Psi =
\log\!\left(
1 +
\left|
\frac{\mathrm{var}(\Delta a)\,\sigma_W}
{m_{\mathrm{eff}}\mu_{|v|} + \epsilon}
\right|
\right),
\end{equation}
    where \(\mathrm{var}(\Delta a)\) is the variance of the first difference of acceleration, \(\sigma_W\) is the standard deviation of the work sequence, \(m_{\mathrm{eff}}\) is the effective mass term, \(\mu_{|v|}\) is the mean absolute velocity, and \(\epsilon\) is a small constant for numerical stability.
\end{itemize}

\subsubsection{Spectral Anchors}
These descriptors capture coarse timbral structure and the frequency-domain footprint of the source.They provide stable spectral anchors for resonance shape, dominant frequency placement, and effective bandwidth, building on standard audio feature families \cite{peeters2004large}. Let \(S(k,t)\) denote the magnitude spectrogram at frequency bin \(k\) and frame \(t\), let \(f_k\) denote the center frequency of bin \(k\), let \(P[k]\) denote the power spectral density, and let \(c[t]\) denote the spectral centroid at frame \(t\).

\begin{itemize}
    \item \textbf{Spectral Anchors} (Spectral Anchor I-III): We include the first three MFCC coefficients as standard low-dimensional summaries of the coarse spectral envelope \cite{davis1980comparison}. In our descriptor set, they serve as compact timbral anchors alongside the proposed acoustically grounded features. Empirically, these three anchors provide strong coarse separation for classes such as \emph{vacuum cleaner}, \emph{thunderstorm}, \emph{airplane}, \emph{church bells}, and \emph{can opening}.

    \item \textbf{Peak Mode} ($f_{\mathrm{mode}}$): This descriptor captures the principal resonant or dominant frequency mode of the source. Lower values are typical of lower-frequency or broadband events such as \emph{footsteps}, \emph{engine}, \emph{helicopter}, and \emph{thunderstorm}, while higher values appear in classes with sharper or higher-frequency dominant modes such as \emph{clock alarm}, \emph{crying baby}, \emph{rooster}, and \emph{frog}. We define it from the dominant spectral peak of the power spectral density as
    \begin{equation}
    f_{\mathrm{mode}} = \log\!\left(1 + \left| f_{\arg\max_k P[k]} \right|\right).
    \end{equation}
    Here, \(P[k]\) denotes the power spectral density at frequency bin \(k\), \(\arg\max_k P[k]\) is the index of the dominant spectral peak, \(f_k\) is the center frequency of bin \(k\), and the outer \(\log(1+|\cdot|)\) term applies log compression.

    \item \textbf{Dissipation Limit} ($\Phi_{\mathrm{dl}}$): This descriptor estimates the upper extent of the effective energy bandwidth. Low $\Phi_{\mathrm{dl}}$ indicates energy concentrated in lower-frequency structure, as in \emph{thunderstorm}, \emph{wind}, \emph{airplane}, and \emph{door knock}. High $\Phi_{\mathrm{dl}}$ indicates broad high-frequency support, as in \emph{brushing teeth}, \emph{pouring water}, \emph{rain}, and \emph{crackling fire}. We compute it as the mean \(85\%\) spectral rolloff frequency:
    \begin{equation}
    \Phi_{\mathrm{dl}} = \mathbb{E}_t[\mathrm{Rolloff}_{85\%}(t)].
    \end{equation}
    Here, \(\mathrm{Rolloff}_{85\%}(t)\) is the spectral rolloff at frame \(t\), i.e., the frequency below which \(85\%\) of the spectral energy is accumulated, and \(\mathbb{E}_t[\cdot]\) denotes the average over time frames.

    \item \textbf{Resonant Spread} ($\Phi_{\mathrm{spread}}$): This descriptor measures how concentrated or dispersed the spectral footprint is around its centroid. Low $\Phi_{\mathrm{spread}}$ is associated with relatively focused spectral structure, as in \emph{church bells}, \emph{siren}, \emph{airplane}, and \emph{door knock}. High $\Phi_{\mathrm{spread}}$ is associated with spectrally scattered or broadband events such as \emph{crackling fire}, \emph{brushing teeth}, \emph{rain}, and \emph{insects}. We compute it as the mean RMS spectral deviation from the centroid:
    \begin{equation}
    \Phi_{\mathrm{spread}} =
    \mathbb{E}_t\!\left[
    \sqrt{
    \frac{\sum_k (f_k - c[t])^2\,|S(k,t)|}
         {\sum_k |S(k,t)| + \epsilon}}
    \right].
    \end{equation}
    Here, \(S(k,t)\) denotes the magnitude spectrogram at frequency bin \(k\) and frame \(t\), \(f_k\) is the center frequency of bin \(k\), \(c[t]\) is the spectral centroid at frame \(t\), \(\mathbb{E}_t[\cdot]\) denotes averaging over time frames, and \(\epsilon\) is a small constant for numerical stability.
\end{itemize}

\subsubsection{Temporal Dynamics}
These descriptors characterize event-scale temporal behavior, focusing on onset sharpness, dissipation regularity, temporal roughness, and the density of salient excitation events. Let \(o[t]\) denote the onset-strength envelope at frame \(t\), let \(o_{\mathrm{seg}}[t]\) denote the same envelope restricted to the initial attack segment, and let \(T\) denote the clip duration.

\begin{itemize}
    \item \textbf{Attack Gradient} ($\nabla_{\mathrm{att}}$): This descriptor captures how the onset envelope evolves immediately after excitation. More negative values correspond to abrupt onset spikes followed by rapid decay, as in \emph{can opening}, \emph{door knock}, and \emph{glass breaking}. Higher values correspond to onset segments that continue to rise or remain active over the initial window, as in \emph{dog}, \emph{sneezing}, and \emph{mouse click}. We compute it as the mean rate of change of the onset-strength envelope over the initial \(50\) ms onset segment:
\begin{equation}
\nabla_{\mathrm{att}} = \mathbb{E}_t\!\left[\nabla o_{\mathrm{seg}}[t]\right].
\end{equation}
Here, \(o_{\mathrm{seg}}[t]\) denotes the segment of the onset-strength envelope beginning at the first frame satisfying \(o[t] > 0.05\max_{\tau}o[\tau]\) and extending for the next \(\left\lfloor 0.05f_s/512 \right\rfloor\) frames, clipped to the available signal length, and \(\nabla\) denotes the discrete gradient operator applied to that segment.

    \item \textbf{Damping Coefficient} ($\zeta$): This descriptor measures the regularity of onset behavior over the full clip. Low \(\zeta\) corresponds to more regular or stable onset behavior, as in \emph{wind}, \emph{airplane}, \emph{sea waves}, and \emph{vacuum cleaner}. High \(\zeta\) indicates irregular or strongly varying excitation and dissipation, as in \emph{glass breaking}, \emph{door knock}, \emph{sneezing}, \emph{coughing}, and \emph{can opening}. We compute it as the coefficient of variation of onset strength:
    \begin{equation}
    \zeta = \frac{\sigma_o}{\mu_o + \epsilon},
    \end{equation}
    where \(\mu_o = \mathbb{E}_t[o[t]]\) is the mean onset strength over time, \(\sigma_o = \mathrm{std}_t(o[t])\) is its standard deviation, and \(\epsilon\) is a small constant for numerical stability.

    \item \textbf{Temporal Jitter} ($\Phi_{\mathrm{jitter}}$): This descriptor measures temporal roughness in the evolution of the onset envelope. Low values indicate smoother energy flow, as in \emph{wind}, \emph{airplane}, and \emph{sea waves}, whereas high values indicate rapidly fluctuating or jagged temporal structure, as in \emph{keyboard typing}, \emph{fireworks}, \emph{mouse click}, and \emph{can opening}. We compute it as the standard deviation of the first difference of the onset-strength sequence:
    \begin{equation}
    \Phi_{\mathrm{jitter}} = \mathrm{std}_t\!\left(\Delta o[t]\right).
    \end{equation}
    Here, \(o[t]\) denotes the onset-strength envelope at frame \(t\), \(\Delta o[t]\) is its first-order difference, and \(\mathrm{std}_t(\cdot)\) denotes the standard deviation across time frames.

    \item \textbf{Pulse Density} ($\Phi_{\mathrm{pulse}}$): This descriptor measures the density of strong salient peaks within the clip. High \(\Phi_{\mathrm{pulse}}\) indicates a large number of strong, repeatedly occurring peaks, as in \emph{clock alarm}, \emph{vacuum cleaner}, \emph{washing machine}, and \emph{crickets}. Low \(\Phi_{\mathrm{pulse}}\) is more typical of clips containing fewer strong peaks, such as \emph{water drops}, \emph{drinking/sipping}, and \emph{footsteps}. We define it as
\begin{equation}
\Phi_{\mathrm{pulse}} = \frac{N_{\mathrm{peaks}}}{T},
\end{equation}
where \(x[n]\) is the normalized audio waveform at sample \(n\), \(N_{\mathrm{peaks}}\) is the number of peaks detected in \(|x[n]|\) with peak height greater than \(0.3\max_n |x[n]|\), and \(T\) is the clip duration.

\end{itemize}

\subsubsection{Stochasticity and Structural Consistency}
These descriptors measure how ordered, irregular, or periodically structured the signal is over time. Let \(S(k,t)\) denote the magnitude spectrogram of the signal at frequency bin \(k\) and frame \(t\), let \(\mathrm{ZCR}[t]\) denote the zero-crossing rate of frame \(t\), and let \(\ell\) denote the autocorrelation lag.

\begin{itemize}
    \item \textbf{Entropy Flow} ($\Phi_{\mathrm{ef}}$): This descriptor measures the average spectral flatness over time \cite{peeters2004large}. Low $\Phi_{\mathrm{ef}}$ indicates more coherent or spectrally structured energy, as in \emph{church bells}, \emph{airplane}, \emph{wind}, and \emph{train}. High $\Phi_{\mathrm{ef}}$ indicates flatter, more broadband, and spectrally disordered events, as in \emph{rooster}, \emph{sneezing}, \emph{coughing}, and \emph{can opening}. We compute it as the mean spectral flatness across frames:
    \begin{equation}
    \Phi_{\mathrm{ef}} = \mathbb{E}_t[\mathrm{Flatness}(S(:,t))].
    \end{equation}
    Here, \(S(:,t)\) denotes the spectral slice of the magnitude spectrogram at frame \(t\), \(\mathrm{Flatness}(\cdot)\) denotes the spectral flatness measure, and \(\mathbb{E}_t[\cdot]\) denotes averaging over time frames.

    \item \textbf{Impact Peakedness} ($\kappa_s$): This descriptor measures how strongly energy concentrates into sharp spectral peaks rather than remaining diffusely distributed. Higher values are associated with signals that contain persistent or pronounced spectral spikes, as in \emph{siren}, \emph{snoring}, \emph{clock alarm}, and \emph{footsteps}, whereas lower values are more typical of flatter or more diffuse spectral structure, as in \emph{sea waves}, \emph{clapping}, \emph{rain}, and \emph{hand saw}. We compute it as the mean spectral kurtosis across frames:
\begin{equation}
\kappa_s = \mathbb{E}_t[\mathrm{Kurtosis}_F(S(:,t))].
\end{equation}
Here, \(S(:,t)\) denotes the spectral slice at frame \(t\), \(\mathrm{Kurtosis}_F(\cdot)\) is the Fisher kurtosis of that slice, and \(\mathbb{E}_t[\cdot]\) denotes averaging over time frames.

    \item \textbf{ZCR Variability} ($\sigma_{\mathrm{zcr}}$): This descriptor measures the variation of zero-crossing behavior over time \cite{peeters2004large}. Low $\sigma_{\mathrm{zcr}}$ indicates more stable fine-scale oscillatory behavior, as in \emph{wind}, \emph{thunderstorm}, \emph{helicopter}, and \emph{airplane}. High $\sigma_{\mathrm{zcr}}$ indicates stronger variation in short-timescale sign changes, as in \emph{glass breaking}, \emph{can opening}, \emph{pouring water}, and \emph{sneezing}. We compute it as the standard deviation of the zero-crossing rate across frames:
    \begin{equation}
    \sigma_{\mathrm{zcr}} = \mathrm{std}_t(\mathrm{ZCR}[t]).
    \end{equation}
    Here, \(\mathrm{ZCR}[t]\) denotes the zero-crossing rate at frame \(t\), and \(\mathrm{std}_t(\cdot)\) denotes the standard deviation across time frames.

    \item \textbf{Periodicity Strength} ($\rho_{\max}$): This descriptor measures how strongly repeating structure is expressed in the signal \cite{peeters2004large}. High $\rho_{\max}$ indicates strong periodicity, as in \emph{clock alarm}, \emph{car horn}, \emph{church bells}, and \emph{engine}. Low $\rho_{\max}$ is characteristic of weakly periodic or aperiodic events such as \emph{sea waves}, \emph{clapping}, \emph{rain}, and \emph{can opening}. We compute it as the maximum normalized autocorrelation excluding the first 50-sample lag:
\begin{equation}
\rho_{\max} = \max_{50 \le \ell < L}\tilde{r}_x[\ell],
\end{equation}
where \(\tilde{r}_x[\ell] = \frac{r_x[\ell]}{r_x[0] + \epsilon}\) is the normalized autocorrelation at lag \(\ell\), \(L\) is the maximum available lag after truncation to \(2000\) samples, and \(\epsilon\) is a small constant for numerical stability.

    \item \textbf{PDE Residual} ($\mathcal{R}_{\mathrm{pde}}$): This descriptor measures the mismatch between temporal acceleration structure and a coarse spectral-curvature estimate. Let \(a[n]\) denote the discrete second-order temporal difference of the waveform and let \(H_{\mathrm{spec}}\) denote the accumulated local spectral curvature around dominant spectral peaks. Low $\mathcal{R}_{\mathrm{pde}}$ indicates signals whose temporal and spectral structure evolve in a comparatively coherent manner, as in \emph{thunderstorm}, \emph{cow}, \emph{door knock}, and \emph{siren}. High $\mathcal{R}_{\mathrm{pde}}$ indicates stronger mismatch between these two views, as in \emph{mouse click}, \emph{can opening}, \emph{crickets}, and \emph{clock alarm}. We define it as
\begin{equation}
H_{\mathrm{spec}} = \sum_{p \in \mathcal{P}}
\left|P[p+1] - 2P[p] + P[p-1]\right|,
\end{equation}
where \(\mathcal{P}\) is the set of up to ten strongest interior peaks of the power spectral density \(P[k]\), detected with minimum peak spacing \(50\) bins and minimum prominence \(0.01\max_k P[k]\).

\begin{equation}
\mathcal{R}_{\mathrm{pde}}
=
\frac{
\left|
\mathbb{E}[a[n]^2] - \frac{H_{\mathrm{spec}}}{f_s^2}
\right|
}{
\mathrm{Var}(x[n]) + \epsilon
}.
\end{equation}
Here, \(a[n]\) is the discrete second-order temporal difference of the waveform, \(H_{\mathrm{spec}}\) is the accumulated local spectral-curvature term defined above, \(f_s\) is the sampling rate, \(\mathrm{Var}(x[n])\) is the variance of the waveform, and \(\epsilon\) is a small constant for numerical stability.

\end{itemize}

\section{Experiments and Results}

\subsection{Datasets}
We evaluate \textit{MADS} on two sound classification benchmarks: the \textit{ESC-50} dataset and its \textit{ESC-10} subset \cite{piczak2015esc}, and the \textit{MSoS} benchmark \cite{kroos2019msos}. Together, these benchmarks allow us to assess the proposed descriptor set across environmental and broader sound classification settings.

\begin{itemize}
    \item \textbf{ESC-50 / ESC-10:} ESC-50 contains 2{,}000 environmental audio clips of 5 seconds each, evenly distributed across 50 classes and organized into five official cross-validation folds. ESC-10 is a 10-class subset of the same corpus and serves as a simpler benchmark for environmental sound classification.
    
\item \textbf{MSoS:} MSoS (Making Sense of Sounds) is a 5-class sound classification benchmark comprising the high-level categories \emph{Effects}, \emph{Human}, \emph{Music}, \emph{Nature}, and \emph{Urban}. We evaluate on MSoS using its official train/test split.
\end{itemize}
\subsection{Feature Extraction and Preprocessing}
For implementation, all extracted features are sanitized by replacing NaN and \(\pm\infty\) values with \(0\), and then clipped to \([-10^6, 10^6]\). In our implementation, \(\epsilon = 10^{-10}\).
For all classical ML experiments, features are z-score standardized using \textit{StandardScaler} fitted on the corresponding training split.

\subsection{Evaluation Protocol and Classifiers}
To evaluate the discriminative utility of the proposed descriptors, we perform sound classification using a standard suite of classical machine learning models. Across all benchmarks, we use 1-NN and 3-NN \cite{cover1967nearest}, logistic regression \cite{cox1958regression}, SVM with an RBF kernel \cite{cortes1995support}, random forest \cite{breiman2001random}, Gaussian Naive Bayes \cite{john1995estimating}, XGBoost \cite{chen2016xgboost}, and a soft-voting ensemble \cite{kittler1998combining}. In Tables~\ref{tab:esc10_compare_multiseed}, \ref{tab:esc50_compare_multiseed}, and~\ref{tab:msos_compare_multiseed}, we report mean classification accuracy and standard deviation over five random seeds (\(42\)--\(46\)).

We compare \textit{MADS} against two conventional handcrafted baselines: \(B_1\), a compact 26D baseline following the ESC setup of Piczak \cite{piczak2015esc}, and \(B_2\), an expanded 38D conventional baseline built from standard MFCC and spectral summary descriptors \cite{davis1980comparison,peeters2004large}. \(B_1\) consists of the mean and standard deviation of zero-crossing rate together with the mean and standard deviation of MFCC coefficients \(1{:}12\), i.e., the \(0^{\mathrm{th}}\) MFCC coefficient is discarded. \(B_2\) expands this setup with a broader set of standard spectral statistics, including MFCC coefficients \(0{:}12\), zero-crossing rate, spectral centroid, spectral bandwidth, spectral rolloff, spectral flatness, and RMS energy. This yields \(13 \times 2 = 26\) MFCC statistics and \(6 \times 2 = 12\) additional summary statistics, for a total of 38 features.

For \textit{ESC-10} and \textit{ESC-50}, we follow the official 5-fold cross-validation protocol independently for each  seed, and then report the mean and standard deviation across the five resulting accuracies. For \textit{MSoS}, we use the official train/test split and report the mean and standard deviation of test accuracy across the same five  seeds.

\subsection{Results and discussion}
\par Tables~\ref{tab:esc10_compare_multiseed}, \ref{tab:esc50_compare_multiseed}, and~\ref{tab:msos_compare_multiseed} show a consistent pattern. First, \textit{MADS} substantially outperforms the compact baseline \(B_1\) across the strongest classifiers on all three benchmarks. On \textit{ESC-10}, \textit{MADS} achieves the best ensemble accuracy (\textbf{81.00\%}) and the best XGBoost accuracy (\textbf{79.60\%}), while also improving over \(B_1\) for Logistic Regression, Naive Bayes, and both k-NN variants. On \textit{ESC-50}, \textit{MADS} again delivers the strongest top-end performance, achieving the best ensemble accuracy (\textbf{52.78\%}), the best Random Forest accuracy (\textbf{49.33\%}), and the best XGBoost accuracy (\textbf{50.13\%}). On \textit{MSoS}, \textit{MADS} likewise achieves the strongest top-end results, reaching the best voting ensemble accuracy (\textbf{67.48\%}), the best Random Forest accuracy (\textbf{67.20\%}), and the best XGBoost accuracy (\textbf{67.28\%}). Second, despite using only 19 features, \textit{MADS} remains highly competitive with the larger conventional baseline \(B_2\) (38D). On \textit{ESC-10}, it outperforms \(B_2\) on the ensemble, XGBoost, Logistic Regression, 3-NN, and Naive Bayes; on \textit{ESC-50}, it remains superior on the best-performing models overall; and on \textit{MSoS}, it exceeds \(B_2\) on the strongest ensemble and tree-based models while using roughly half the dimensionality.

\begin{figure}[t]
    \centering
    \hspace*{-0.04\linewidth}
    \includegraphics[width=1.18\linewidth]{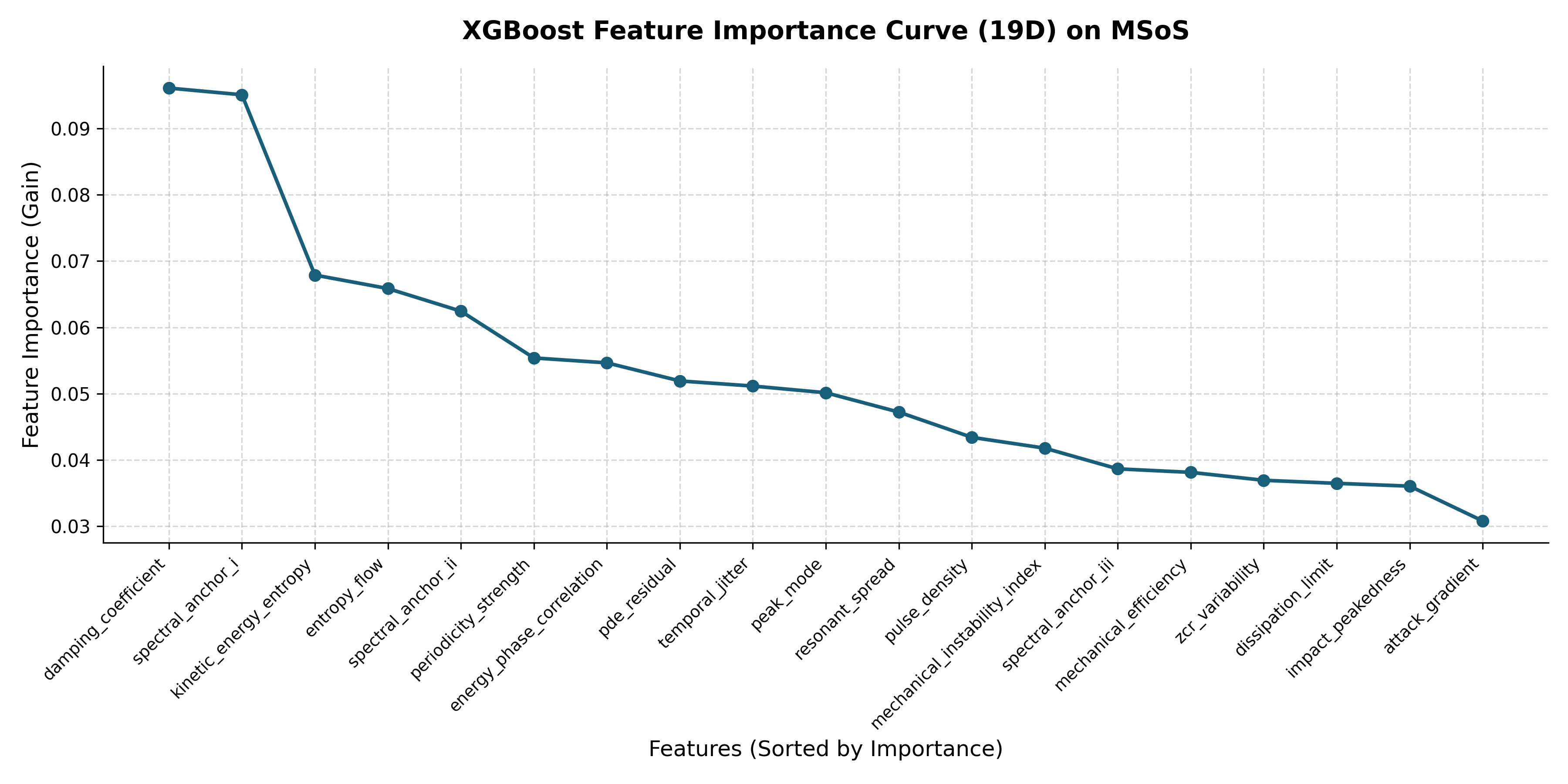}
    \caption{XGBoost feature-importance curve (gain) for \textit{MADS} on \textit{MSoS}. The importance profile spans multiple descriptor families, indicating that the classifier uses complementary temporal, spectral, mechanical, and stochastic cues.}
    \label{fig:msos_xgb_importance}
\end{figure}
A further view of the proposed representation is provided by the XGBoost feature-importance analysis on \textit{MSoS}, shown in Fig.~\ref{fig:msos_xgb_importance}. The learned importance profile is not concentrated on a single descriptor family; instead, high-gain features are drawn from temporal, spectral, mechanical, and stochastic components of \textit{MADS}. In particular, descriptors such as \textit{damping coefficient}, \textit{spectral anchor i}, \textit{kinetic energy entropy}, and \textit{entropy flow} emerge among the most influential variables. This suggests that the classifier is exploiting the intended multi-view structure of \textit{MADS}, rather than relying only on a narrow subset of conventional spectral cues.

\begin{table}[t]
\centering
\caption{\textit{ESC-10} 5-fold cross-validation comparison. Values are mean accuracy (\%) over 5 seeds (\(42\)--\(46\)).}
\label{tab:esc10_compare_multiseed}
\small
\setlength{\tabcolsep}{4pt}
\begin{tabular}{lccc}
\toprule
Model & \(B_1\) (26D) & \(B_2\) (38D) & Ours (19D) \\
\midrule
Voting (RF+XGB+LR) & 74.00 $\pm$ 0.50 & 79.05 $\pm$ 0.57 & \textbf{81.00 $\pm$ 0.25} \\
Random Forest      & 72.45 $\pm$ 0.84 & \textbf{79.45 $\pm$ 0.41} & 78.70 $\pm$ 0.62 \\
XGBoost            & 70.95 $\pm$ 0.82 & 79.20 $\pm$ 0.37 & \textbf{79.60 $\pm$ 0.52} \\
SVM (RBF)          & 70.25 $\pm$ 0.00 & \textbf{77.25 $\pm$ 0.00} & 73.25 $\pm$ 0.00 \\
Logistic Regression & 68.25 $\pm$ 0.00 & 72.00 $\pm$ 0.00 & \textbf{75.50 $\pm$ 0.00} \\
3-NN               & 65.50 $\pm$ 0.00 & 68.25 $\pm$ 0.00 & \textbf{68.50 $\pm$ 0.00} \\
Naive Bayes        & 65.50 $\pm$ 0.00 & 70.50 $\pm$ 0.00 & \textbf{72.25 $\pm$ 0.00} \\
1-NN               & 64.00 $\pm$ 0.00 & \textbf{69.00 $\pm$ 0.00} & 66.25 $\pm$ 0.00 \\
\bottomrule
\end{tabular}
\end{table}

\begin{table}[t]
\centering
\caption{\textit{ESC-50} 5-fold cross-validation comparison. Values are mean accuracy (\%) over 5 seeds (\(42\)--\(46\)).}
\label{tab:esc50_compare_multiseed}
\small
\setlength{\tabcolsep}{4pt}
\begin{tabular}{lccc}
\toprule
Model & \(B_1\) (26D) & \(B_2\) (38D) & Ours (19D) \\
\midrule
Voting (RF+XGB+LR) & 42.73 $\pm$ 0.24 & 51.78 $\pm$ 0.21 & \textbf{52.78 $\pm$ 0.33} \\
SVM (RBF)          & 41.75 $\pm$ 0.00 & \textbf{47.50 $\pm$ 0.00} & 41.95 $\pm$ 0.00 \\
Random Forest      & 41.59 $\pm$ 0.60 & 47.20 $\pm$ 0.31 & \textbf{49.33 $\pm$ 0.62} \\
XGBoost            & 39.47 $\pm$ 0.45 & 46.16 $\pm$ 0.34 & \textbf{50.13 $\pm$ 0.19} \\
Logistic Regression & 39.35 $\pm$ 0.00 & \textbf{46.40 $\pm$ 0.00} & 45.25 $\pm$ 0.00 \\
Naive Bayes        & 33.50 $\pm$ 0.00 & \textbf{35.65 $\pm$ 0.00} & 33.40 $\pm$ 0.00 \\
1-NN               & 33.05 $\pm$ 0.00 & \textbf{35.65 $\pm$ 0.00} & 34.15 $\pm$ 0.00 \\
3-NN               & 32.15 $\pm$ 0.00 & \textbf{35.65 $\pm$ 0.00} & 31.60 $\pm$ 0.00 \\
\bottomrule
\end{tabular}
\end{table}

\begin{table}[t]
\centering
\caption{\textit{MSoS} classification comparison. Values are mean accuracy (\%) over 5  seeds (\(42\)--\(46\)) evaluated on the official test set.}
\label{tab:msos_compare_multiseed}
\small
\setlength{\tabcolsep}{4pt}
\begin{tabular}{lccc}
\toprule
Model & \(B_1\) (26D) & \(B_2\) (38D) & Ours (19D) \\
\midrule
Voting (RF+XGB+LR)   & 60.80 $\pm$ 1.00 & 66.04 $\pm$ 0.41 & \textbf{67.48 $\pm$ 0.41} \\
Random Forest        & 60.88 $\pm$ 0.73 & 63.12 $\pm$ 0.95 & \textbf{67.20 $\pm$ 0.66} \\
XGBoost              & 61.84 $\pm$ 0.82 & 66.84 $\pm$ 0.41 & \textbf{67.28 $\pm$ 1.14} \\
Logistic Regression  & 48.40 $\pm$ 0.00 & \textbf{53.60 $\pm$ 0.00} & 51.60 $\pm$ 0.00 \\
Naive Bayes          & \textbf{45.60 $\pm$ 0.00} & 44.40 $\pm$ 0.00 & 42.80 $\pm$ 0.00 \\
1-NN                 & 59.80 $\pm$ 0.00 & \textbf{65.80 $\pm$ 0.00} & 62.20 $\pm$ 0.00 \\
3-NN                 & 58.80 $\pm$ 0.00 & \textbf{63.00 $\pm$ 0.00} & 60.40 $\pm$ 0.00 \\
\bottomrule
\end{tabular}
\end{table}

\section{Conclusions}\label{s:con}

In this work, we introduced \textit{MADS}, a compact 19-dimensional physics-informed descriptor set for sound classification. \textit{MADS} was designed to capture complementary spectral, temporal, mechanical, and stochastic properties of an audio signal within a unified multi-view representation, motivated by the hypothesis that acoustically grounded structure can provide a stronger and more interpretable basis for recognition than conventional low-order spectral summaries alone.

Across \textit{ESC-10}, \textit{ESC-50}, and \textit{MSoS}, the proposed descriptors consistently outperformed the compact \(B_1\) handcrafted baseline and achieved the strongest peak results overall on the best-performing models, while remaining competitive with a substantially larger \(B_2\) conventional feature set despite using only 19 dimensions. These results show that carefully designed acoustically grounded descriptors can recover discriminative structure that is not well captured by standard spectral summary features alone.

More broadly, we view the present work as foundational rather than terminal. Beyond its value as a compact standalone representation, \textit{MADS} defines a structured descriptor layer that can support future frame-level and deep-learning-compatible extensions. Exploring such temporal lifts of the descriptor set, and studying how acoustically grounded representations can interact with modern neural audio models, remains an important direction for follow-up work.
\subsection{Future Directions}
While the present work focuses on validating \textit{MADS} as a compact 19-dimensional descriptor set, the proposed representation is not intended to terminate at a clip-level summary. Many of the descriptors introduced here are computed from underlying temporal trajectories, spectral evolutions, and mechanically motivated intermediate quantities. \textit{MADS} should therefore be understood as the first compact layer of a richer acoustically grounded representation space.

A natural next step is to lift these descriptors from clip-level aggregates into frame-level or segment-level tracks, yielding a structured temporal representation in which each channel corresponds to a physically meaningful acoustic view, such as excitation strength, damping behavior, periodicity, stochasticity, or spectral anchoring. Such a representation would differ fundamentally from conventional log-mel inputs: rather than exposing only a generic time-frequency energy pattern, it would explicitly organize the signal according to interpretable acoustic dynamics.

This direction opens a path toward deep-learning-compatible extensions of \textit{MADS}, including sequence models and hybrid architectures that operate on structured concept channels rather than only on fixed spectral frontends. In this sense, the present paper lays the descriptor foundation for a broader representation program: one that seeks to combine compact acoustically grounded structure, interpretability, and compatibility with modern temporal modeling.

\section*{Declaration of Competing Interest}

\par The authors have no relevant financial or non-financial interests to disclose.

\bibliography{sn-bibliography}

\end{document}